\documentclass[11pt]{article}
\usepackage{amsmath}
\usepackage{graphicx}
\usepackage{amssymb}
\usepackage{amsfonts}
\usepackage{latexsym}

\renewcommand{\baselinestretch}{1.1}
\def\beq{\begin{eqnarray}}
\def\eeq{\end{eqnarray}}

\def\al{\alpha}
\def\be{\beta}

\def\ga{\gamma}
\def\de{\delta}

\def\ka{\kappa}
\def\la{\lambda}
\def\na{\nabla}

\def\si{\sigma}
\def\om{\omega}

\def\ta{\tau}
\def\th{\theta}

\def\Om{\Omega}

\begin{document}
\vskip 0.7cm

\begin{center}

{\large\bf Fourth derivative gravity in the auxiliary fields representation
and application to the black hole stability}
\vskip 4mm

{\bf Sebasti\~{a}o Mauro}$^{a}$,
\ \
{\bf Roberto Balbinot}$^{b,c,d}$,
\ \
{\bf Alessandro Fabbri}$^{d,b,e,f}$,
\\
{\bf and}
\ \
{\bf Ilya L. Shapiro}$^{a,g}$
\vskip 4mm

{\it (a)} \ \ Departamento de F\'{\i}sica, ICE,
Universidade Federal de Juiz de Fora,
\\ CEP: 36036-330, Juiz de Fora,
MG, Brazil
\vskip 2mm

{\it (b)} \ \
Dipartimento di Fisica dellUniversit\`{a} di Bologna,
Via Irnerio 46,
\\
I-40126 Bologna, Italy
\vskip 2mm

{\it (c)} \ \
INFN Sezione di Bologna, Via Irnerio 46, I-40126 Bologna, Italy
\vskip 2mm

{\it (d)} \ \
Centro Studi e Ricerche E. Fermi,
Piazza del Viminale 1, I-00184 Roma, Italy
\vskip 2mm

{\it (e)} \ \
Departamento de Fisica Teorica and IFIC, Universidad de Valencia-CSIC,
\\ C. Dr. Moliner, 60, I-46100 Burjassot, Spain
\vskip 2mm

{\it (f)} \ \
Laboratoire de Physique Th\'{e}orique d'Orsay, B\^{a}timent
210, Universit\'{e} Paris-Sud 11, F-91405, Orsay Cedex, France
\vskip 2mm

{\it (g)} \ \
Tomsk State Pedagogical University and Tomsk State
University,
\\
Tomsk, 634041, Russia
\end{center}
\vskip 8mm

\begin{quotation}
\noindent
{\bf Abstract.}
\ \
We consider an auxiliary fields formulation for the general fourth-order
gravity on an arbitrary curved background. The case of a Ricci-flat
background is elaborated in full details and it is shown that there is
an equivalence with the standard metric formulation. At the same time,
using auxiliary fields helps to make perturbations to look simpler and
the results more clear. As an application we reconsider the linear
perturbations for the classical Schwarzschild solution. We also briefly
discuss the relation to the effect of massive unphysical ghosts in the
theory.
\vskip 4mm

\noindent
Pacs: 04.62.+v, \ 	
04.30.Nk,       \   
04.70.Bw        \   
\vskip 2mm

\end{quotation}

\section{Introduction}

The long-standing interest to higher derivative gravity theories is
based on their critical importance for quantum extensions of gravity.
In this case the general fourth order gravity represents a minimal
UV completion in the vacuum (purely metric) sector of the theory.
Such a theory is renormalizable at both semiclassical \cite{UtDW}
(see also \cite{birdav,book} for the introduction) and quantum
\cite{Stelle-77} levels. At the same time the higher derivative
quantum gravity can not be considered as a completely consistent
theory, because its spectrum of states includes unphysical massive
ghost - the spin-2 state with negative kinetic energy. Starting
from 70-ies there were several different proposals on how to
solve the problem of massive ghosts
\cite{Stelle-77,Tomboulis-77,salstr,antomb},
mainly based on the quantum field theory methods, but all of them
did not lead to conclusive results \cite{johnston}. Very recently
one of us proposed a more simple classical way of dealing with
ghosts \cite{HD-Stab}.
The new approach to the problem is based on the study of
stability of the classical solutions in the low-energy sector of
the theory, and in this respect it is similar to the consideration
based on the effective quantum field theory ideas \cite{Burgess}.

The linear stability of the given solution of some differential
equation is usually sufficient for
the stability at any perturbative level. Therefore, in order to
observe the effect of the spin-2 ghosts one can simply look at
the asymptotic behaviour of the gravitational waves on the given
background. In the case of cosmological solutions discussed in
\cite{HD-Stab}, we could observe that the ghost does not produce
instability below the threshold, which approximately corresponds
to the Planck scale of frequencies. Only starting from the Planck
scale one can observe an explosive-type dynamics of the
gravitational waves. The situation is completely different for
tachyons, in this case an instability takes place independent
on the energy scale. This important difference was recently 
discussed in \cite{CreNic} and \cite{GWprT}.

The stability of the cosmological background due
to the huge mass of the ghost \cite{HD-Stab} is in accordance
with the previously known result for the particular deSitter
background \cite{star83,wave,HHR}. In these papers the
anomaly-induced semiclassical contributions were taken into
account, but finally were shown to be irrelevant \cite{GW-Stab}.
It would be definitely interesting to consider other physically
relevant background solutions, in particular for the Schwarzschild
metric.
During some time the unique study of the stability of the
$d=4$ classical Schwarzschild solution in the presence of
higher derivative gravitational terms was the work by Whitt
\cite{whitt}. The main result is that the solution is stable
in the case of fourth derivative gravity. Recently there was
another paper by Myung, \cite{myung}, which argued in favor of
the opposite result, by using the analogy with the well-known
Gregory-Laflamme instabilities \cite{gregory} (see also
\cite{fabbri}). The difference between
\cite{whitt} and \cite{myung} does not concern the equations
for perturbations, but only their solutions.

In this short report we discuss the procedure of separating
Ricci-flat and higher-derivative perturbations of \cite{whitt}.
For this end we employ the method of auxiliary fields, as it
was described in a recent paper \cite{fReddi} (see also
\cite{faraoni} for standard reviews and further references).
In our opinion, the possibility to present the general fourth
derivative gravity in the second-order form by means of
tensor auxiliary fields has some independent interest. On
the top of that we apply this new description to the
analysis of metric perturbations on the background of
Schwarzschild solution.

\section{Standard approach to fourth-derivative perturbations}

In this section we summarize the Whitt approach \cite{whitt}, which
will be verified in the next section by using auxiliary fields.
According to \cite{whitt},  it is possible to explore the stability
by means of the second-order equations for the perturbation of the
Ricci tensor, instead of the fourth order equations in the metric
perturbations. This approach helps to simplify the standard
treatment of the stability problem in the theory of the fourth
order in derivatives.

Let us consider the
gravitational action of the fourth order theory,
\beq
S[g_{\mu \nu}]
&=&
\int d^4x \sqrt{-g} \Big(
- \frac{1}{\ka ^2}\, R - \frac{\om}{3\la}\, R^2
- \frac{1}{2\la} \,C_{\mu\nu\al\be}^2
\Big)\,,
\label{8a}
\eeq
where $C_{\mu\nu\al\be}$ is the Weyl tensor,
$C_{\mu\nu\al\be}^2 = E + 2(R_{\mu\nu}^2-1/3\,R^2)$. In this
expression we use notations which are typical in quantum gravity,
in particular the transition to the notations of Whitt \cite{whitt}
can be performed as follows:
\beq
16 \pi G = \ka^2
\,,\quad
\frac{1}{\la} = \frac{\be}{16 \pi G}
\,,\quad
\frac{1-\om}{3\la} = \frac{\al}{16 \pi G}\,.
\label{albe}
\eeq
Furthermore, we disregard the Gauss-Bonnet term with the integrand
$E=R_{\mu\nu\al\be}^2-4R_{\mu\nu}^2+R^2$, since it is not relevant
at the classical level.

The variation of the action (\ref{8a}) with respect to the
metric yields
\beq
&&
\frac{1}{\ka^2}\,\Big( R_{\mu\nu} - \frac12\,R g_{\mu\nu}\Big)
+ g_{\mu\nu}\Big[
\frac{1-\om}{6\la}\,R^2
- \frac{1}{2\la}\,R^2_{\rho\si}
+ \frac{4\om-1}{6\la}\, \Box R \Big]
\nonumber
\\
&&
- \frac{2\om + 1}{3\la}\, \na_\mu \na_\nu R
+ \frac{1}{\la}\,\Box R_{\mu\nu}
+ \frac{2}{\la}\, R^{\rho\si}\,R_{\rho\mu\si\nu}
+ \frac{2(\om - 1)}{3\la}\, R\, R_{\mu\nu}
\,=\, 0\,.
\label{2nn}
\eeq
Taking trace we arrive at
\beq
\frac{1}{\ka^2}\,R +
\frac{2\om}{\la}\,\Box R \,=\, 0 \,.
\label{pert5compl}
\eeq
According to this equation, the scalar curvature obeys the
covariant minimal Klein-Gordon equation. Furthermore, in case
of $\om=0$ the last formula boils down to the non-dynamical
constraint on the scalar curvature $R=0$. The choice
$\om=0$ corresponds to the case of a pure Weyl-squared higher
derivative part in Eq. (\ref{8a}). The higher derivative part
of the action possess local conformal invariance, the symmetry
which is violated only by the Einstein-Hilbert term.

Equation (\ref{2nn}) shows that any vacuum solution of Einstein
equations is also a solution for the theory of the fourth order
(\ref{8a}). For the perturbations this may be not true, but in
the case when the background metric satisfies vacuum Einstein
equations, they can be classified into the two distinct classes:

\begin{itemize}

\item
The perturbations  which satisfy Einstein's equations in
vacuum, $R_{\mu \nu}(g_{\alpha \beta} + h_{\alpha \beta})=0$.

\item
The ones which do not satisfy these equations.
\end{itemize}

In the first case we have exactly the same problem as in
Einstein's theory because
\beq
R_{\mu \nu}(g_{\al \be}
+ h_{\al \be})= R_{\mu \nu}(g_{\al \be})
+ R_{\mu\nu}^{(1)}( h_{\al \be})
= R_{\mu\nu}^{(1)}( h_{\al \be})=0\,,
\label{3}
\eeq
where $R_{\mu\nu}^{(1)}$ is the first order expansion of the Ricci
tensor, which will be specified below.

This ensures, for example, the stability of the Schwarzschild
metric, as it was discussed in \cite{RW,Zer,Vis}. In the second
case one can consider the perturbation of the Ricci tensor,
\beq
R^{\prime}_{\mu \nu} = R_{\mu \nu} + R_{\mu\nu}^{(1)}\,.
\label{4}
\eeq
Substituting (\ref{4}) into (\ref{2nn}) and remembering that
$R_{\mu \nu}(g_{\alpha \beta})=0$, we obtain the equation for
the linear perturbations of the Ricci tensor
\beq
&&
\frac{1}{\ka^2}\,
\Big[ R^{(1)}_{\mu\nu} - \frac12\, g_{\mu\nu} R^{(1)}\Big]
+ \frac{4\om-1}{6\la}\, g_{\mu\nu}\Box R^{(1)}
- \frac{2\om+1}{6\la}\, \na_\mu\na_\nu R^{(1)}
\nonumber
\\
&& + \frac{1}{\la}\, \Box R^{(1)}_{\mu\nu}
+ \frac{2}{\la}\, R^{(1)\rho\si}\, R_{\mu\si\nu\rho}
\,=\, 0\,.
\label{5}
\eeq
where we use notation $R^{(1)} = g_{\mu \nu} R_{\mu\nu}^{(1)}$.
Let us note that in the Ricci-flat case $R^{(1)}$ is also the
first order term in the expansion of the curvature of the
second type.

Taking the trace of (\ref{5}) we get
\beq
\frac{1}{\ka^2}\,R^{(1)} +
\frac{2\om}{\la}\,\Box R^{(1)} = 0 \,,
\label{pert5}
\eeq
which is the linearized form of Eq. (\ref{pert5compl}). For
$\om=0$ this gives the constraint $\,R^{(1)}=0$.

In what follows we shall limit our analysis to
this particular case\footnote{For different values of $\om$,
the trace free part and the trace part of $R^{(1)}_{\mu\nu}$
do not decouple leading to a fourth order equation for
$\tilde R_{\mu\nu}^{(1)}$, where
$\tilde R^{(1)}_{\mu\nu}
= R_{\mu\nu}^{(1)}-\frac{1}{4}g_{\mu\nu}R^{(1)}$.}.
So eq. (\ref{5}) becomes
\beq
\Box R^{(1)}_{\mu\nu}
+ 2R_{\tau\mu\lambda\nu}R^{(1)\tau\lambda}
+ \frac{\la}{\ka^2}R^{(1)}_{\mu\nu}\,=\,0\,.
\label{n1}
\eeq
Eq. (\ref{n1}) determines the dynamics of small gravitational
perturbations. It turns out that the stability of the background
Schwarzschild metrics in the fourth order theory in $d=4$ is
defined by the same equations as in the black string case in
$d=5$ space-time dimensions \cite{gregory} and as in the
bi-metric theory of gravity \cite{fabbri}. As it was noted
in \cite{myung}, in these cases one meets a well-known
Gregory-Laflamme instability \cite{gregory}.

\section{Using auxiliary fields}

Let us consider the action with auxiliary fields $\,\phi_{\mu\nu}$,
\beq
S_2
&=&
\int d^4x \sqrt{-g}
\Big\{- \frac{1}{\ka ^2} R + \phi_{\mu\nu}R^{\mu\nu}
+ \frac{\xi}{2}\,\phi_{\mu\nu} 
(A^{-1})^{\mu\nu , \al\be} \phi_{\al\be}\Big\}\,,
\label{9}
\eeq
where $A^{\mu\nu , \al\be}$ is an invertible $c$-number operator
depending only on the metric. The variation of this action with
respect to the field $\phi_{\mu\nu}$ gives
\beq
\phi_{\al \be} &=& - \,\frac{1}{\xi}\, A_{\al\be,\mu\nu}R^{\mu\nu} \,.
\label{11}
\eeq
After using the last relation, the action $S_2$ on-shell becomes
\beq
S_{2}
&=&
\int d^4x \sqrt{-g} \Big(- \frac{1}{\ka ^2} R
- \frac{1}{2\xi}\, R^{\mu\nu} A_{\mu\nu ,\al\be} R^{\al\be}\Big) \,.
\label{12}
\eeq
Comparing (\ref{8a}) with (\ref{12}) we obtain
\beq
\xi &=& \frac{\la}{2}
\,,
\qquad
A_{\mu\nu ,\al\be}
\,=\,
\de_{\mu\nu ,\al\be} - \ga g_{\mu\nu} g_{\al\be}\,,
\label{13}
\eeq
where $\ga = (1-\om)/3$. Here we use the standard DeWitt notation
\beq
\de_{\mu\nu ,\al\be}
&=& \frac{1}{2}(g_{\mu\al} g_{\nu \be} + g_{\mu\be} g_{\nu \al}) \,.
\label{14}
\eeq
The operator defined by (\ref{13}) does not have an inverse if
$\ga = 1/4$, that is for $\om = 1/4$. Let us simply
assume\footnote{In the special case $\ga = 1/4$ the
formulation via auxiliary tensor field is possible, but this
field must be traceless. It is not difficult to elaborate this
version, but there is no much practical interest to do it.}
$\,\ga \neq 1/4$.
Then the inverse operator can be easily obtained
with
\beq
A^{-1}_{\mu\nu , \al\be} \,\, A^{\al\be ,\ta\la}
\,=\,
\de_{\mu\nu ,}^{\,\,\,\,\,\,\, \ta \la}  
\eeq
and one can check that
\beq
A^{-1}_{\mu\nu ,\al\be}
&=& \de_{\mu\nu ,\al\be} - \th g_{\mu\nu} g_{\al \be}\,,
\label{16}
\eeq
where
\beq
\th = \frac{1 - \om}{1-4\om} \, . \quad
\label{17}
\eeq
In this way we arrive at the explicit form of Eq. (\ref{9}),
\beq
S_{2} &=&  \int d^4x \sqrt{-g}
\Big\{
- \frac{1}{\ka ^2} R + \phi_{\mu\nu}R^{\mu\nu}
+ \frac{\la}{4}\, \phi_{\mu\nu} (\de^{\mu\nu , \al\be}
- \th g^{\mu\nu} g^{\al \be}) \phi_{\al \be}  \Big\} \,.
\label{18}
\eeq
It is natural to check whether the auxiliary field $\phi_{\al \be}$
satisfied some constraints. First of all, without losing generality
we can assume that $\phi_{\al \be}$ is a symmetric space-time tensor.
Furthermore, one can use Eqs. (\ref{11}) and (\ref{13}) to derive the
Bianchi identity for this field in the form
\beq
\na_\mu \phi^{\mu}_{\,\,\nu}
&=&
\Big(\frac12 - \th\Big)\na_\nu \phi\,,
\quad
\mbox{where}
\quad
\phi = \phi^{\mu\nu} g_{\mu\nu}\,.
\label{Bi}
\eeq
There is a complicated question on whether the auxiliary field
should satisfy the constraint (\ref{Bi}) at the quantum level, but
since in the present work our attention is restricted to the
purely classical case, this relation should be respected.

The action (\ref{18}) describes a dynamical theory of the two
symmetric tensor fields $g_{\mu\nu}$ and $\phi_{\mu\nu}$. It is
important to note that the auxiliary field is dynamical, because
variation with respect to the metric produce second derivatives
of $\phi_{\mu\nu}$ in the equations of motion. The action
(\ref{18}) is second-derivative, but it is dynamically equivalent
to the original fourth-derivative action (\ref{8a}). This
equivalence means one can map any solution of one theory to
some solution of another one and vice versa. Indeed, this is true
for both background and perturbed solutions, so the action (\ref{18})
can be useful to explore the perturbations.

In order to illustrate how it works, let us expand this action
up to the second order
\beq
S_2 = S_2^{(0)} + S_2^{(1)} + S_2^{(2)}
\label{19}
\eeq
in the perturbations of both metric and auxiliary field.
The expansions of the fields are defined as
\beq
g_{\mu \nu} \to \tilde{g}_{\mu \nu}
&=& g_{\mu\nu} + h_{\mu\nu} \,,
\label{20a}
\\
\phi_{\mu \nu} \to \tilde{\phi}_{\mu \nu}
&=& \phi_{\mu\nu} + \psi_{\mu\nu} \,.
\label{20b}
\eeq
In what follows we shall also use notations for the traces
$\,h=h^\mu_\mu=h_{\mu\nu}g^{\mu\nu}$ and  $\,\psi = \psi^\mu_\mu$.
One can easily check that the first terms in the expansions
$R^{(1)}_{\mu\nu}$ and $\psi_{\mu\nu}$ satisfy the same
Bianchi identities as the background quantities,
\beq
\na_\mu R^{(1)\mu}_\nu
\,=\,\frac12\,\na_\nu R^{(1)}\,,
\qquad
\na_\mu \psi^{\mu}_\nu
\,=\,\Big(\frac12-\th\Big)\,\na_\nu \psi\,.
\label{Bi-aux}
\eeq

Other relevant quantities are expanded to the second order
as follows:
\beq
\tilde{g}^{\mu \nu} &=& g^{\mu\nu} - h^{\mu\nu}
+ h^{\mu\la} h_{\la}^{\nu}\, +\, \dots\,,
\nonumber
\\
\sqrt{-\tilde{g}} &=& \sqrt{-g} \big(1 + \frac{1}{2} h
- \frac{1}{4} h_{\al\be} h^{\al\be} + \frac{1}{8} h^2
\, +\, \dots \big)\,.
\nonumber
\eeq
Furthermore, the first-order expansion for the Ricci tensor is
\beq
R^{(1)}_{\mu\nu}
&=&
-\frac{1}{2} \Box h_{\mu\nu}
- \frac{1}{2} \na_\mu \na_\nu h
+ \frac{1}{2} \na_\la \na_\mu h^{\la}_{\nu}
+ \frac{1}{2} \na_\la \na_\nu h^{\la}_{\mu}
\label{21}
\\
&=&
\frac{1}{2}\,\Big(
  \na_\mu\na_\la h^\la_\nu
+ \na_\nu\na_\la h^\la_\mu
- \Box h_{\mu\nu}
- \nabla_\mu\nabla_\nu h
+ R_\mu^\la h_{\la\nu}+ R_\nu^\la h_{\la\mu}
-2R_{\alpha\mu\beta\nu}h^{\alpha\beta}
\Big)
\nonumber
\eeq
and the second-order term is
\beq
R^{(2)}_{\mu\nu}
&=&
\frac{1}{2} h^{\ta \la}(\na_\ta \na_\la h_{\mu\nu}
+ \na_\mu \na_\nu h_{\ta \la} - \na_\la \na_\nu h_{\ta \mu}
- \na_\la \na_\mu h_{\ta \nu } )
\nonumber
\\
&+&
\frac{1}{2} \big(\na_\ta h^{\ta \la} - \frac{1}{2} \na^\la h\big)
\big(\na_\la h_{\mu\nu} - \na_\mu h_{\la \nu }
- \na_\nu h_{\la \mu }\big)
\nonumber
\\
&+&
\frac{1}{4} \na_\nu h^{\ta \la} \na_\mu h_{\ta \la}
+ \frac{1}{2}\na_\ta h_{\la \mu} \na^\ta h^{\la}_{\nu}
- \frac{1}{2} \na_\ta h^{\la}_{\mu} \na_\la h^{\ta}_{\nu} \,.
\label{22}
\eeq

The Ricci scalar in the first order is
\beq
R^{(1)} &=& \na_\mu \na_\nu h^{\mu\nu} - \Box h
- R_{\mu\nu} h^{\mu\nu} \ ,
\label{23}
\eeq
and in the second-order we meet
\beq
R^{(2)}
&=&
h^{\mu\nu} \big( \Box h_{\mu\nu} + \na_\mu \na_\nu h
- \na_\mu \na_\la h^{\la}_{\nu}
- \na_\la \na_\mu h^{\la}_{\nu}\big)
+ \na_\mu h \na_\nu h^{\mu\nu}
- \na_\mu h^{\mu\nu} \na_\la h^{\la}_{\nu}
\nonumber
\\
 &-&
  \frac{1}{4} \na_\la h \na^\la h
 + \frac{3}{4} \na_\la h_{\mu\nu} \na^\la h^{\mu\nu}
- \frac{1}{2} \na_\la h_{\mu\nu} \na^\mu h^{\la\nu}
+ R_{\mu\nu} h^{\mu\la}h^{\nu}_{\la} \ .
\label{24}
\eeq
Replacing these formulas into the action $S_2$ (\ref{18}),
we obtain the first-order expansion
\beq
S^{(1)}_2
&=&
\int d^4x \sqrt{-g} \Big\{
R^{\mu\nu} \psi_{\mu\nu}
\,+\,\frac{h}{2}\big[
- \frac{1}{\ka ^2} R
+ \phi_{\mu\nu}R^{\mu\nu}
+ \frac{\la}{4}
(\phi_{\mu\nu}\phi^{\mu\nu} - \th \phi^2) \big]
\nonumber
\\
&-& 2\phi_{\mu\al} R^\al_\nu h^{\mu\nu}
- \frac{\la}{2}\, \phi_{\mu\al} \phi^\al_\nu h^{\mu\nu}
+ \frac{\la}{2}\, \th \phi \phi_{\mu\nu} h^{\mu\nu}
+ \frac{\la}{2}
(\phi^{\mu\nu} \psi_{\mu\nu} - \th \phi \psi)
\nonumber
\\
&-&
\frac{1}{\ka^2} \big( \na_\mu \na_\nu h^{\mu\nu}
- \Box h - R_{\mu\nu}h^{\mu\nu}\big)
+ \phi^{\mu\nu} \big(
\na_\la \na_\mu h^{\la}_{\nu}
- \frac{1}{2} \na_\mu \na_\nu h - \frac{1}{2} \Box h_{\mu\nu}
\big)
\Big\} \,.
\label{25}
\eeq
Finally, the second-order term is
\beq
S^{(2)}_2
&=&
\int d^4x \sqrt{-g} \,
\Big\{ \frac{\la}{4}\,(\phi_{\mu\nu}\phi^{\mu\nu} - \th \phi^2)
- 2 \psi^\mu_\be R_{\mu\nu} h^{\be\nu}
- \la \phi_{\mu\nu} \psi^\mu_\be h^{\be\nu}
+ \phi^{\mu\nu} R^{(2)}_{\mu\nu}
\nonumber
\\
&+&
\frac{\la \th}{2}\,
\big(\phi \psi_{\al\be} h^{\al\be} + \psi \phi_{\al\be} h^{\al\be}\big)
+ \frac{h}{2}
\big[ \psi^{\mu\nu}R_{\mu\nu}
+ \frac{\la}{2} (\phi^{\mu\nu} \psi_{\mu\nu} - \th \phi \psi)
+ \phi^{\mu\nu} R^{(1)}_{\mu\nu} \big]
\nonumber
\\
&+&
\big( \psi^{\mu\nu} - 2\phi^\mu_\be\, h^{\nu\be}\big)R^{(1)}_{\mu\nu}
+ \big(\phi_{\al\be} h^{\mu\al}h^{\nu\be}
+ 2 \phi^\mu_\be h^{\be\la} h^\nu_\la
-  \phi^\mu_\be  h^{\nu\be} h\big)R_{\mu\nu}
\nonumber
\\
&-&
\frac{\la}{4}\,h\big(
\phi_{\mu\nu}\phi^\mu_\be h^{\nu\be}
- \th\phi\phi_{\al\be} h^{\al\be} \big)
- \frac{1}{\ka^2}\,\big[R^{(2)} + \frac{1}{2}\,h R^{(1)} \big]
\nonumber
\\
&+&
\phi_{\mu\nu}\phi_{\al\be}\big[\frac{\la}{4} h^{\mu\al}h^{\nu\be}
+ \frac{\la}{2}g^{\mu\al} h^{\nu\la}h_{\la}^{\be}
- \frac{\la\th}{4} h^{\mu\nu}h^{\al\be}
- \frac{\la\th}{2}g^{\mu\nu} h^{\al\la}h_{\la}{\be}\big]
\nonumber
\\
&+&
\big(\frac{1}{8} h^2 -\frac{1}{4} h_{\al\be} h^{\al\be} \big)
\big[\phi^{\mu\nu}R_{\mu\nu}
+ \frac{\la}{4}\,(\phi_{\mu\nu}\phi^{\mu\nu} - \th \phi^2)
- \frac{1}{\ka^2}\,R\,\big]
\Big\} \ .
\label{26}
\eeq

The expression (\ref{26}) contains the complete equations for the
perturbations. However, for the sake of simplicity we assume that 
the background metric satisfies Einstein equations in vacuum, 
such that we take $R_{\mu\nu}=0$ and $\phi_{\mu\nu}=0$. After 
discarding surface terms, we obtain the simplified expression
\beq
S^{(2)}_2
&=&
\int d^4x \sqrt{-g} \Big\{
\frac{\la}{4}\, \psi_{\mu\nu} \psi_{\al\be}
(\de^{\mu\nu , \al\be} - \th g^{\mu\nu} g^{\al \be})
\nonumber
\\
&+&
\psi^{\mu\nu}
\Big[-\frac{1}{2} \Box h_{\mu\nu} - \frac{1}{2} \na_\mu \na_\nu h
+   \na_\mu \na_\la h^{\la}_{\nu}
- R_{\mu\al\nu\be} h^{\al\be} \Big]
\nonumber
\\
&+&
\frac{1}{2\ka^2}\,h^{\mu\nu}\, \Big[\frac{1}{2}g_{\mu\nu}\Box h
+ \na_\mu \na_\la h^{\la}_{\nu} -\frac{1}{2} \Box h_{\mu\nu}
- g_{\mu\nu} \na_\al \na_\be h^{\al\be}
- R_{\mu\al\nu\be} h^{\al\be} \Big] \Big\} \,.
\label{27}
\eeq
The equations that determine the dynamics of the perturbations
have the form
\beq
-\, \frac{\la}{2}\,  (\de_{\mu\nu ,\al\be}
- \th g_{\mu\nu} g_{\al \be})\psi^{\al\be}
\,=\,R^{(1)}_{\mu\nu}
\label{28}
\eeq
and
\beq
&&
\frac{1}{\ka^2}\,\big(
  g_{\mu\nu} \Box h
- g_{\mu\nu} \na_\al \na_\be h^{\al\be}
+ \na_\mu \na_\la h^{\la}_{\nu}
+ \na_\nu \na_\la h^{\la}_{\mu}
- \na_{\mu}\na_{\nu}h
- \Box h_{\mu\nu}
-  2R_{\mu\al\nu\be} h^{\al\be}\big)
\nonumber
\\
&&
+\, \na_\mu \na_\la \psi^{\la}_{\nu}
+ \na_\nu \na_\la \psi^{\la}_{\mu}
- \Box \psi_{\mu\nu}
- g_{\mu\nu} \na_\al \na_\be \psi^{\al\be}
- 2R_{\mu\al\nu\be} \psi^{\al\be}
\,=\, 0 \,.
\label{29-pre}
\eeq
By means of relations (\ref{Bi-aux}) and (\ref{23}) the last equation
can be presented in the form
\beq
&&
\frac{1}{\ka^2}\,\big[
R^{(1)}_{\mu\nu} - \frac12\,g_{\mu\nu}\,R^{(1)}\big]
\,=\,
\frac12\, \Box \psi_{\mu\nu}
+ R_{\mu\al\nu\be} \psi^{\al\be}
+ \Big(\th-\frac12\Big)
\Big(\na_\mu \na_\nu \psi - \frac12\,g_{\mu\nu} \Box\psi \Big)\,.
\label{29}
\eeq

Inverting Eq. (\ref{28}) and inserting it into Eq. (\ref{29}) one
can reproduce Eq. (\ref{5}). For the conformal fourth-derivative
case, when $\om=0$, the equation is
\beq
\label{a}
\Box R^{(1)}_{\mu\nu}
+ 2R_{\tau\mu\lambda\nu}R^{(1)\tau\lambda}
+ \frac{\la}{\ka^2}R^{(1)}_{\mu\nu}\,=\,0\,,
\eeq
where we used the constraint $R^{(1)}=0$ (consequently $\psi=0$ too). Furthermore, one can impose the $TT$-gauge on the metric fluctuations,
\beq
h=0\,,
\quad
\nabla_\mu h^{\mu\nu}=0\,,
\label{TT}
\eeq
so that Eq. (\ref{21}) which relates metric and Ricci fluctuations
becomes
\beq
\label{b}
\Box h_{\mu\nu}+2R_{\tau\mu\la\nu}h^{\tau\la}
+ 2R^{(1)}_{\mu\nu}=0\,,
\eeq
or, in terms of the auxiliary field,
\beq
\Box h_{\mu\nu}+2R_{\tau\mu\lambda\nu}h^{\tau\lambda}
&=&
\la\,\psi_{\mu\nu}\,.
\label{32}
\eeq
In the last formula we took into account the constraint $\psi=0$ which
holds in the case $\om=0$. The same constraint, together with the TT-gauge
(\ref{TT}), reduce the Eq. (\ref{29-pre}) to the relation
\beq
\Box h_{\mu\nu} +  2R_{\mu\al\nu\be} h^{\al\be}
\,=\,
-\,\ka^2\,\big(\Box \psi_{\mu\nu}
+ 2R_{\mu\al\nu\be} \psi^{\al\be}\big)\,.
\label{29-TT}
\eeq
Taken together, Eqs. (\ref{32}) and (\ref{29-TT}) lead to the massive
equation for the auxiliary field alone,
\beq
\Box \psi_{\mu\nu}
+ 2R_{\mu\al\nu\be} \psi^{\al\be}
\,+\,
\frac{\la}{\ka^2}\, \psi_{\mu\nu}\,=\,0\,.
\label{29-m}
\eeq

\section{Brief discussion}

In the previous section we have shown that the two descriptions
of perturbations are indeed equivalent. Let us see how one can
apply the auxiliary fields description for the black-hole
background case.

Starting from equations (\ref{a}) and (\ref{b}) one can define a
new field
\beq
\si_{\mu\nu}
&=&
\frac{1}{\ka^2}\,h_{\mu\nu} + \psi_{\mu\nu}
\,=\,
\frac{1}{\ka^2}\,h_{\mu\nu} - \frac{2}{\la}\,R^{(1)}_{\mu\nu}\,,
\label{sigma}
\eeq
which satisfies the massless equation (on a Ricci-flat background)
\beq
\Box \si_{\mu\nu} + 2R_{\mu\al\nu\be} \si^{\al\be} = 0 \,.
\label{38}
\eeq
This is exactly the equation for the Ricci-flat perturbations in
Einstein theory, which were explored by Regge, Wheeler et al in
\cite{RW,Zer,Vis}. It is known to have no unstable modes.
Therefore, as far as unstable modes are concerned we can
set $\sigma_{\mu\nu}=0$, i.e.
$\,\psi_{\mu\nu} = - h_{\mu\nu}/\ka^2$,
making Eq. (\ref{b}) to take the same form as Eq. (\ref{a}).

In order to establish the relation between massive ghosts and
instabilities, one has to note that the ghost with the mass of the
Planck order of magnitude does not produce instabilities in flat
space or in a very weak gravitational field \cite{CreNic,Burgess}.
The practical consequence for the black hole case should be a
selection rule for the boundary conditions of the perturbations
at space infinity.
Neglecting the oscillatory part, we can impose the requirement for
those perturbations which are related to the ghost instabilities that
$\psi_{\mu\nu},\ h_{\mu\nu}$ vanish for large values of r, where the
background solution approaches Minkowski space.
Only those solutions for the perturbations should
be permitted, if we intend to estimate the effect of massive
ghost on the instabilities on the black-hole background.

The stability of the perturbations is completely defined by eq.
(\ref{a}). This equation is known to possess a spherically symmetric
unstable mode provided
\beq
0 < \frac{1}{\sqrt{\beta}} = \frac{\sqrt{\la}}{\ka} < \frac{O(1)}{r_S}\,,
\nonumber
\eeq
where $r_S$ is Schwarzschild radius \cite{gregory}.
Such a mode grows exponentially in time as $e^{\Omega t}$, $\Omega>0$,
spatially vanishes at infinity and is regular at the future horizon.
This mode is present in both $R^{(1)}_{\mu\nu}$ (as also noted in
\cite{myung}) and $h_{\mu\nu}$.

The Gregory-Laflamme unstable mode exponentially decays in the radial
direction, leading to
\beq
\big(h_{\mu\nu},\,
\psi_{\mu\nu}\big)
\,\sim\, e^{\Om t - r \sqrt{\Omega^2 + \la/\ka^2}}\,.
\label{exp}
\eeq
Eq. (\ref{exp}) tells us that in any given finite space point at some
instant of time
the perturbations become large and then show an unrestricted growth.
According to our previous considerations, one can consider this as an
indication that the Gregory Laflamme instability is not related to the
presence of massive ghosts in the spectrum. Finally, we note that the
end-state of this instability is thought to be a black hole with a
massive graviton halo
\footnote{In the context of massive gravity a
numerical analysis \cite{bcp} showed that for graviton mass
$\,\sim 1/r_S\,$ likely candidates are black holes with
massive graviton halo, but none have been found for smaller masses.}.

\section{Conclusions}
\label{s6}

We presented a new form of the action and of equations of motion
for the higher derivative gravity, which uses auxiliary fields on
an arbitrary curved space-time background. Previously, the same
representation has been known in flat space \cite{Stelle78} and
for the special de\,Sitter background \cite{HT}. The main result
is the eq. (\ref{26}) from which one can easily extract the full
set of equations in the physically interesting cases or
approximations. For the sake of simplicity, we elaborate only
the perturbations for the Ricci-flat background and arrive the
coupled set of equations for metric and auxiliary field
perturbations.

As an application we consider linear perturbations around the
static spherically symmetric black hole. The use of auxiliary
tensor field enables one to distinguish the perturbations
related to the massive unphysical ghost which is present in
the spectrum of the theory. As a result we can see that these
perturbations can not be identified with the Gregory-Laflamme
instability.

\section*{Acknowledgements}

A.F. thanks E. Babichev for useful discussion.
I.Sh. thanks G. Cusin for helpful conversation
concerning references.
S.M. is grateful to FAPEMIG and CAPES for partial support.
I.Sh. is grateful to CNPq, FAPEMIG and ICTP for partial
support of his work.

\renewcommand{\baselinestretch}{0.9}


\end{document}